\documentclass[a4paper,twoside]{article}

\usepackage{graphicx}
\usepackage{subcaption}
\usepackage{calc}
\usepackage{amssymb}
\usepackage{amstext}
\usepackage{amsmath}
\usepackage{amsthm}
\usepackage{multicol}
\usepackage{pslatex}
\usepackage{apalike}
\usepackage{SCITEPRESS}     % Please add other packages that you may need BEFORE the SCITEPRESS.sty package.

\usepackage[table]{xcolor}

\usepackage[final]{microtype}

\usepackage{hyperref}
\usepackage{cleveref}

\definecolor{codebg}{rgb}{0.98,0.98,0.86}
\begin{document}

\title{SimFaaS: A Performance Simulator for Serverless Computing Platforms}

\author{\authorname{Nima Mahmoudi\sup{1}\orcidAuthor{0000-0002-2592-9559} and Hamzeh Khazaei\sup{2}\orcidAuthor{0000-0001-5439-8024}}
\affiliation{\sup{1}Electrical and Computer Engineering, University of Alberta, Edmonton, Alberta, Canada}
\affiliation{\sup{2}Electrical Engineering and Computer Science, York University, Toronto, Ontario, Canada}
\email{nmahmoud@ualberta.ca, hkh@yorku.ca}
}

% \author{\authorname{Anonymous Authors}
% }


\abstract{Developing accurate and extendable performance models for serverless
platforms, aka Function-as-a-Service (FaaS) platforms, is a very challenging task. Also, implementation and
experimentation on real serverless platforms is both costly and
time-consuming. However, at the moment, there is no comprehensive
simulation tool or framework to be used instead of the real platform.
As a result, in this paper, we fill this gap by proposing a simulation platform, called
SimFaaS, which assists serverless application developers 
to develop optimized Function-as-a-Service applications in terms of cost and performance. 
On the other hand, SimFaaS can be leveraged by FaaS providers to tailor their platforms to be workload-aware so that they can increase profit and quality of service at the same time. 
Also, serverless platform providers can evaluate new designs, implementations, and deployments 
on SimFaaS in a timely and cost-efficient manner. 

SimFaaS is open-source, well-documented, and publicly available, making it easily 
usable and extendable to incorporate more use case scenarios in the future.
Besides, it provides performance engineers
with a set of tools that can calculate several characteristics of
serverless platform internal states, which is otherwise hard
(mostly impossible) to extract from real platforms.
In previous studies, temporal and steady-state performance
models for serverless computing platforms have been developed. However, those models are limited
to Markovian processes. We designed SimFaaS as a tool that can help
overcome such limitations for performance and cost prediction in serverless
computing.

We show how SimFaaS facilitates the prediction of
essential performance metrics such as average response time, probability of
cold start, and the average number of instances reflecting
the infrastructure cost incurred by the serverless computing provider.
We evaluate the accuracy and applicability of SimFaaS by comparing the
prediction results with real-world traces from Amazon AWS Lambda.
}

\onecolumn \maketitle \normalsize \setcounter{footnote}{0} \vfill

\section{INTRODUCTION} \label{sec:intro}

\noindent
% FaaS Scheduling
There is very little official documentation made publicly available about the scheduling algorithms
in public serverless computing platforms. However, many works have focused on partially reverse engineering
this information through experimentations on these 
platforms~\cite{wang2018peeking,figiela2018performance,lloyd2018serverless}.
Using the results of such studies and by modifying their code base and thorough extensive experimentation, we have come
to a good understanding of the way modern serverless frameworks are operated and managed by the service providers.
In this work, we plan to use this information to build an open and public performance 
simulator\footnote{A simulator focusing on the performance-related key metrics and aspects of the system.} for modern serverless computing platforms with a high degree
of flexibility, fidelity and accuracy.

In serverless computing platforms, computation is done in function instances. These instances are
completely managed by the serverless computing platform provider and act as tiny servers for
the incoming triggers (requests). To develop a comprehensive simulator for serverless
computing platforms, we first need to understand how they work underneath and are managed.

The simulator presented in this work is written in \textit{Python}.
The resulting package can easily be installed using 
pip\footnote{
\url{https://pypi.org/project/simfaas/}
}.
The source code is openly accessible on the project
Github\footnote{
\url{https://github.com/pacslab/simfaas}
}. The documentation is accessible on 
Read the Docs\footnote{
\url{https://simfaas.readthedocs.io/en/latest/}
}.
For more information, interested readers can check
out our Github repository, which provides links to all of our artifacts as
well as easy-to-setup environments, to try out our sample scenarios.

The remainder of the paper is organized as follows:
\Cref{sec:sys-desc} describes the system simulated in SimFaaS
in detail. \Cref{sec:simfaas-design} outlines the design of
SimFaaS with the most important design choices and characteristics.
\Cref{sec:sample-use} lists some of possible use cases
for SimFaaS. In \Cref{sec:exp-eval}, we present the
experimental evaluation of SimFaaS, validating the
accuracy of the simulator. \Cref{sec:related-work} gives a
summary of the related work. Finally, \Cref{sec:conc}
concludes the paper.

\section{SYSTEM DESCRIPTION} \label{sec:sys-desc}

\noindent
In this section, we introduce the management system in serverless computing
platforms, which has been fully captured by the serverless simulator presented in
this paper.

\textbf{Function Instance States:} % initializing/running/idle
according to recent
studies~\cite{mahmoudi2020tccserverless,mahmoudi2020tempperf,wang2018peeking,figiela2018performance,mahmoudi2019optimizing},
we identify three states for each function instance: \textit{initializing}, \textit{running}, and \textit{idle}.
The \textit{initializing} state happens when the infrastructure is spinning up new instances, which might include
setting up new virtual machines, unikernels, or containers to handle the excessive
workload. The instance will remain in the \textit{initializing} state until it is able to handle incoming requests.
As defined in this work, we also consider \textit{application initializing}, which is the time user's code is
performing initial tasks like creating database connections, importing libraries or
loading a machine learning model from an S3 bucket as a part of
the \textit{initializing} state which needs to happen
only once for each new instance. Note that the instance cannot accept incoming requests
before performing all initialization tasks. It might be worth noting that the \textit{application initializing} state
is billed by most providers while the rest of the \textit{initializing} state is not billed.
When a request is submitted to the instance, the instance goes into the \textit{running} state. In this state,
the request is parsed and processed. The time spent in the \textit{running} state is also billed by the serverless
provider. After the processing of a request is over, the serverless platform keeps the instances warm
for some time to be able to handle later spikes in the workload. In this state, we consider the
instance to be in the \textit{idle} state. The application developer is not charged for an instance that is in the \textit{idle} state.

\textbf{Cold/Warm start:}
as defined in previous work~\cite{lloyd2018serverless,wang2018peeking,figiela2018performance},
we refer to a request as a \textit{cold start} request when it goes through the process of launching a new
function instance. For the platform, this could include launching a new virtual machine, deploying a
new function, or creating a new instance on an existing virtual machine, which introduces an
overhead to the response time experienced by users. In case the platform has an instance
in the \textit{idle} state when a new request arrives, it will reuse the existing function instance instead
of spinning up a new one. This is commonly known as a \textit{warm start} request. Cold starts could be orders of magnitude longer than warm starts for some applications.
Thus, too many cold starts could impact the application's responsiveness 
and user experience~\cite{wang2018peeking}. This is the reason a lot of research in the field of serverless
computing has focused on mitigating cold starts~\cite{lin2019mitigating,bermbachusing,manner2018cold}.

\textbf{Autoscaling:} % scale-per-request/other type
we have identified three main autoscaling patterns among the mainstream
serverless computing platforms: 1) \textit{scale-per-request}; 2) 
\textit{concurrency value scaling}; 3) \textit{metrics-based scaling}.
In \textit{scale-per-request} Function-as-a-Service (FaaS) platforms, when a request comes in, it will
be serviced by one of the available idle instances (\textit{warm start}), or the platform will spin up a
new instance for that request (\textit{cold start}). Thus, there is
no queuing involved in the system, and each cold start causes the creation
of a new instance, which acts as a tiny server for subsequent requests.
As the load decreases, to scale the number of instances down,
the platform also needs to scale the number of instances down.
In the \textit{scale-per-request} pattern,
as long as requests that are
being made to the instance are less than the \textit{expiration threshold} apart, the instance will be kept warm. In other words, for each instance,
at any moment in time, if a request has not been received in the last \textit{expiration threshold} units of time, it will be
expired and thus terminated by the platform, and the consumed resources will be released. 
To enable simplified billing, 
most well-known public serverless computing platforms
use this scaling pattern, e.g., AWS Lambda, Google Cloud Functions,
IBM Cloud Functions, Apache OpenWhisk, and Azure Functions~\cite{wang2018peeking,van2018spec}.
As scale-per-request is the dominant scaling technique used by major providers, in this paper, we strive to simulate the performance of this type of serverless platform. 

% ~\cite{gcpconcurrency}
\begin{figure}[htbp]
\centerline{\includegraphics[width=0.99\columnwidth]{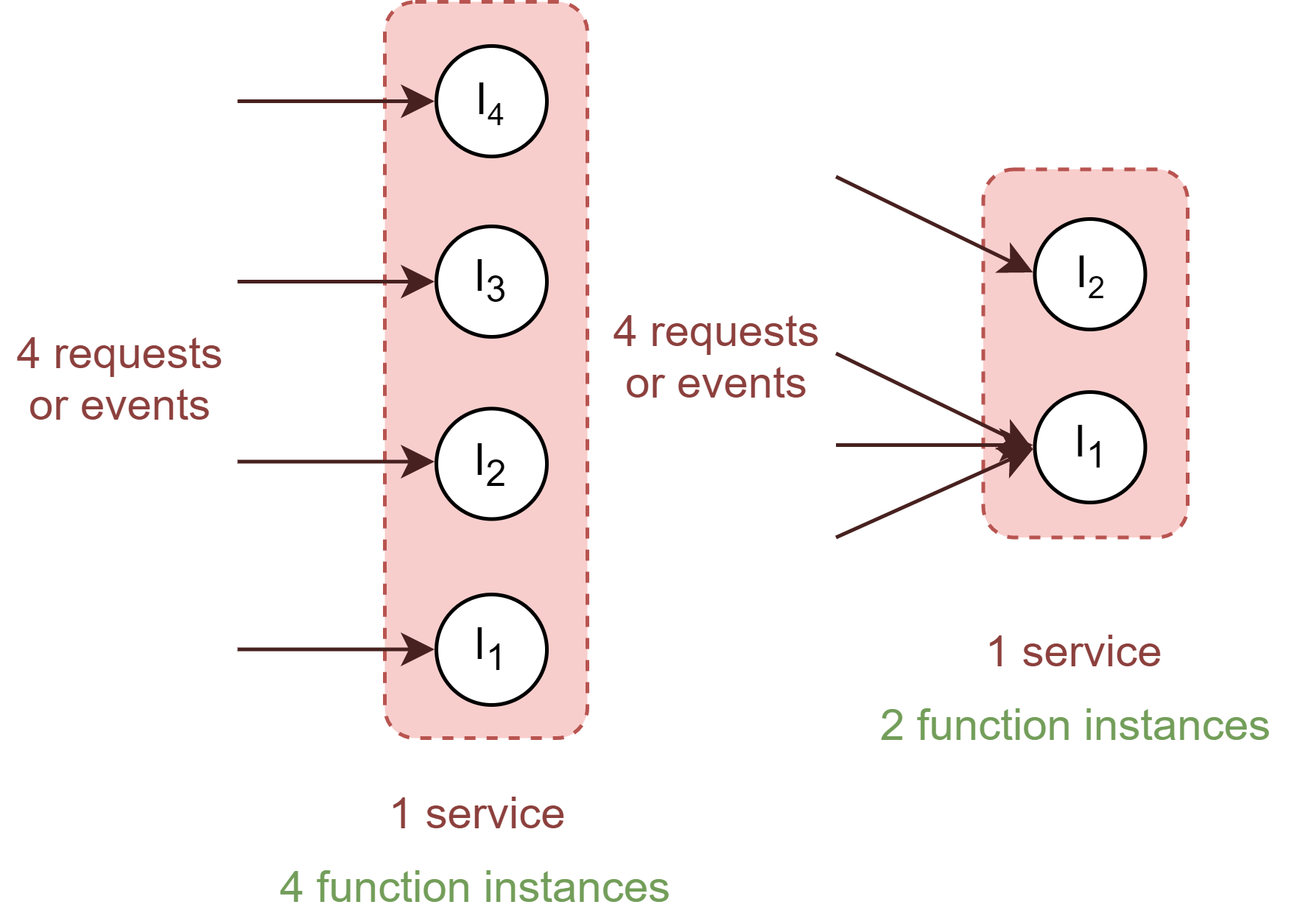}}
\caption{The effect of the concurrency value on the number of function instances needed. The left service has a concurrency value of 1, while the right service has a concurrency value of 3.}
\label{fig:concurrency-value}
\end{figure}

% https://cloud.google.com/run/docs/about-concurrency
In the \textit{concurrency value scaling} pattern~\cite{gcpconcurrency},
function instances can receive multiple requests at the same time.
The number of requests that can be made concurrently to the same instance
can be set via \textit{concurrency value}. \Cref{fig:concurrency-value}
shows the effect of concurrency value on the autoscaling behaviour
of the platform. It is worth noting that the scale-per-request autoscaling
pattern can initially be seen as a special case of \textit{concurrency value scaling} pattern where
concurrency value is set to 1. However, due to its popularity, importance, 
and fundamental differences in their management layer, we classify them into separate categories.
Examples of this scaling pattern are Google Cloud Run and Knative.

\textit{Metrics-based scaling} tries to keep metrics like CPU or memory
usage within a predefined range. Most on-premises serverless computing
platforms work with this pattern due to its simplicity and reliability.
Some of the serverless computing platforms that use this pattern are
AWS Fargate, Azure Container Instances, OpenFaaS, Kubeless, and Fission.

The simulator proposed in this work considers only the platforms
that use the scale-per-request pattern due to their importance and
widespread adoption in mainstream public serverless computing platforms.

\textbf{Initialization Time:}
as mentioned earlier, when the platform is spinning up new instances, they will first go into
the initialization state.
% This state might include spinning up new virtual machines, unikernels, or containers.
The initialization time is the amount of time it takes since the platform
receives a request until the new instance is up and running and ready to serve the request.
The initialization time, as defined here, is comprised of the platform initialization time
and the application initialization time. The platform initialization time is the time it takes for
the platform to make the function instance ready, whether a unikernel or a container and the application initialization time is the time it takes for the application to run the initialization code, e.g., connecting to the database.

\textbf{Response Time:}
the response time usually includes the queuing time and the service time.
Since we are addressing the scale-per-request serverless computing platforms here, 
there is no queuing involved for the incoming requests.
Due to the inherent linear scalability in serverless computing
platforms~\cite{lloyd2018serverless,wang2018peeking,figiela2018performance},
the distribution of the response time does not change over time with different loads.

\textbf{Maximum Concurrency Level:}
every public serverless computing platform has some limitations on the number of function instances that can be
spun up and in running state for a single function. This is mainly due to ensuring the availability of the service for others,
limiting the number of instances one user can have up and running at the same time. This is mostly known
as the \textit{maximum concurrency level}. For example, the default maximum concurrency level for AWS Lambda
is 1000 function instances in 2020 for most regions. When the system reaches the maximum concurrency level, any request
that needs to be served by a new instance will receive an error status showing the server is not
able to fulfill that request at the moment. 

\textbf{Request Routing:}
in order to minimize the number of containers that are kept warm and thus to free up system resources,
the platform routes incoming requests to new containers, and it will use older containers only if all
containers that are created more recently are busy~\cite{mcgrath2017serverless}. 
In other words, the scheduler gives priority to newly instantiated idle instances using priority scheduling according to creation time, i.e., the newer the instance, the higher the priority.
By adopting this
approach, the system minimizes the number of requests going to older containers, maximizing
their chance of being expired and terminated.

\section{THE DESIGN OF SIMFAAS} \label{sec:simfaas-design}

% \begin{figure*}[!ht]
% \centering
% \includegraphics[width=\linewidth]{figs/06-uml-class-diagram.pdf}
% \caption{The UML class diagram of SimFaaS.}
% % generated by pynsource
% \label{fig:06-uml-class}
% \end{figure*}

\begin{figure}[!ht]
\centering
\includegraphics[width=\linewidth]{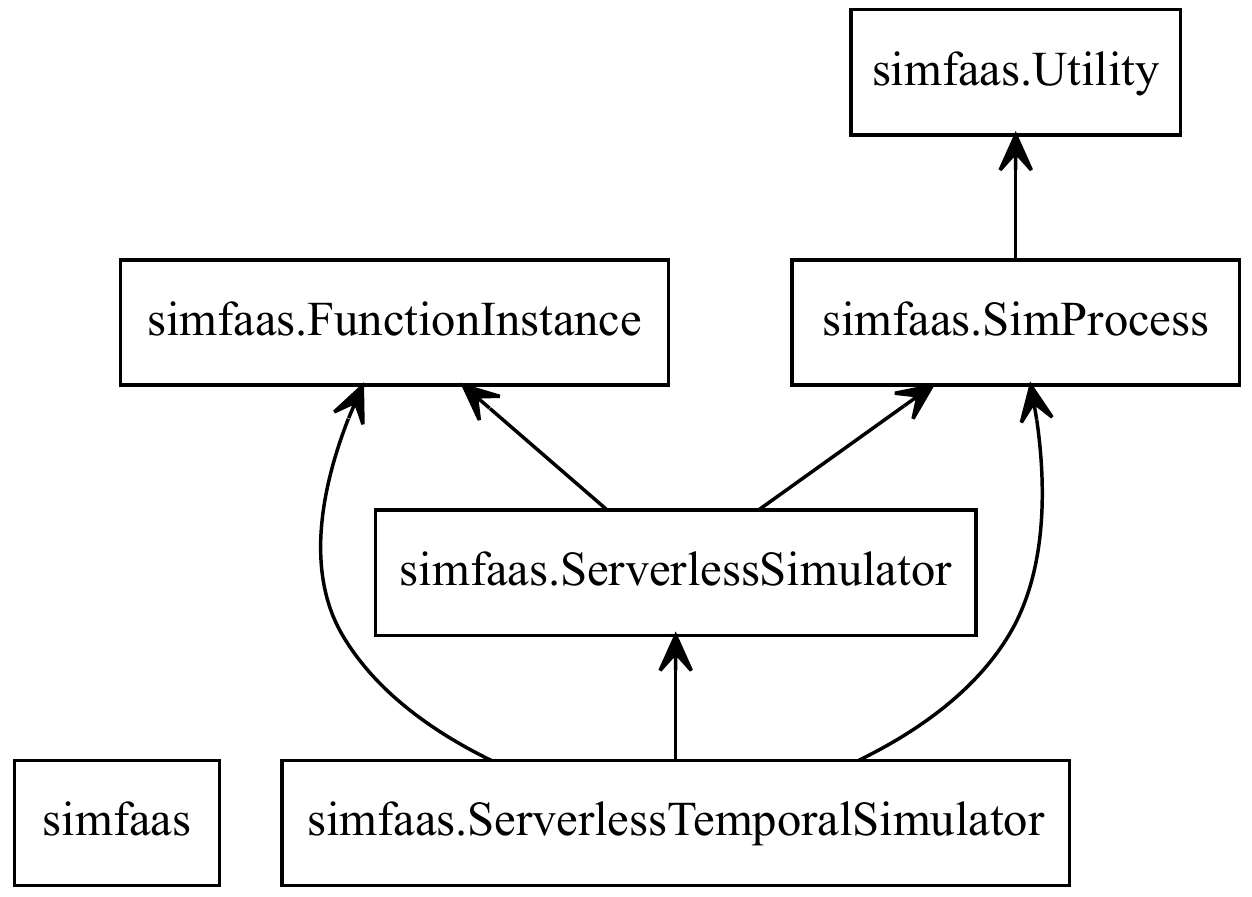}
\caption{The package diagram of SimFaaS. Each square represents a class in the package and arrows represent dependency.}
% generated by pyreverse
\label{fig:package-diagram}
\end{figure}

\noindent
This section discusses the design of the novel Function-as-a-Service
(FaaS) platform simulator (SimFaaS) proposed in this work.
SimFaaS was created by the authors as a tool for simplifying
the process of validating a developed performance model 
and allowing accurate performance prediction for providers
and application developers in the absence of one.
SimFaaS mainly targets public serverless computing platforms. There are several
built-in tools for visualizing, analyzing, and verifying a
developed analytical performance model. In addition, we added
tools that can accept custom state encoding and generate approximations for
Probability Density Functions (PDF) and Cumulative Distribution
Functions (CDF) from the simulations, which can help debug several
parts of a given analytical performance model.

The proposed simulator can predict several QoS-related metrics accurately like cold start
probability, average response time, and the probability of rejection for requests under
different load intensities, which helps application developers understand the limits of
their system and measure their SLA compliance without the need for expensive experiments.
In addition, it can predict the average number of running server count and total server count,
which helps predict the cost of service for the application developer and the infrastructure cost
incurred by the serverless provider, respectively.

% \Cref{fig:06-uml-class} shows the UML class diagram of SimFaaS.
\Cref{fig:package-diagram} outlines the package diagram of
SimFaaS, showing the dependency between different modules.
The \textit{Utility} module provides helper functions for plots and calculations.
The \textit{SimProcess} module will help simulate a single process and allows
for comparisons with the optional analytical model provided as a function handle
to the module. The \textit{FunctionInstance} module provides the functionality of a
single function instance. The main simulation with an empty initial state is provided
by the \textit{ServerlessSimulator} module. Finally, \textit{ServerlessTemporalSimulator}
module performs simulations similar to \textit{ServerlessSimulator} module, but with added
functionality allowing customized initial state and calculation of simulation results in
a time-bounded fashion.

% \begin{table}[h]
% \caption{The functionality provided by each module in SimFaaS.}\label{tab:modules} \centering
% \centering
% \rowcolors{2}{white}{gray!25}
% \begin{tabular}{p{1.8cm}|p{4.8cm}}
%   \hline
%   \textbf{Module} & \textbf{Functionality} \\
%   \hline
%   Serverless-Simulator & Main performance simulation functionality without custom initial state \\
%   \hline
%   SimProcess & Simulating a process and comparison against analytical estimate \\
%   \hline
%   Utility & Utility functions for plotting and calculations \\
%   \hline
% \end{tabular}
% \end{table}

\subsection{Extensibility and Ease of Use}

SimFaaS has been developed entirely in Python using an
object-oriented design methodology. In order to leverage the
tools within the package, the user needs to write a Python
application or a Jupyter notebook initializing the classes
and providing the input parameters. In addition, the user has
the option to extend the functionality in the package by
extending the classes and adding their custom functionality.
Almost every functionality of classes can be overridden
to allow for modification and extensions. For example, the
arrival, cold start service, and warm start service processes
can be redefined by simply extending the \textit{SimProcess}
class.
We included deterministic, Gaussian, and Exponential processes
as examples of such extensions in the package.
Examples of such changes can be found in the several
examples we have provided for SimFaaS.
In addition, the user can include their analytically produced 
PDF and CDF functions to be compared against the simulation
trace results.

The simulator provides all of the functionality needed for
modelling modern scale-per-request serverless computing platforms.
However, we created a modular framework that can span future
types of computational platforms. To demonstrate this, we extended the
\textit{ServerlessSimulator} class to create
\textit{ParServerlessSimulator}, which simulates serverless platforms
that allow queuing in the function instances but have a scaling algorithm
similar to scale-per-request platforms.

\subsection{Support for Popular Serverless Platforms}

SimFaaS includes simulation models able to mimic the most popular
public serverless computing platforms like AWS Lambda, Google
Cloud Functions, IBM Cloud Functions, Apache OpenWhisk,
Azure Functions, and all other platforms
with similar autoscaling. We have also performed over one month
of experimentation to demonstrate the validity of
the simulation results extracted from SimFaaS.

% \subsubsection{Scale-Per-Request Simulation}
% Scale-Per-Request scaling pattern is the most widespread scaling
% pattern in public serverless computing platforms, which emphasizes
% its importance for building a serverless platform simulator.
% AWS Lambda, which is the most widespread serverless computing
% platform, uses this as its scaling pattern. 

To capture the
exogenous parameters needed for an accurate simulation, the
following information is needed:

\begin{itemize}
    \item \textbf{Expiration Threshold} which is usually constant for any given public serverless computing platform. According
    to our experimentations and other works~\cite{shahrad2020serverless,shilkovaws10mins},
    in 2020,
    this value is 10 minutes for AWS Lambda, Google Cloud Functions,
    IBM Cloud, and Apache OpenWhisk, and 20 minutes for
    Azure Functions. For other serverless computing platforms,
    experimentation is needed by the users. The use of a
    non-deterministic expiration threshold is also possible
    by extending the \textit{FunctionInstance} class.
    \item \textbf{The arrival process} which can rather safely be
    assumed as an exponential for most consumer-facing applications.
    However, other applications might use a deterministic process,
    e.g. cron jobs, or other types like batch arrival. The user
    can use one of our built-in processes or simply define their
    own.
    \item \textbf{The warm/cold service process} can be extracted
    by measuring the response time from monitoring the workload
    response time for cold and warm requests. By default, SimFaaS uses exponential distribution
    for this process but can be overridden by the user by
    passing any class that extends the \textit{SimProcess} class. We have provided
    Gaussian and fixed-interval distributions as part of the package to demonstrate this.
\end{itemize}

% \subsection{Built-in Data Collection}

%%%%%%%%%%%%%%%%%%%%%%%%%%%%%%%%%%%%%%%%%%%%%%%%%%%%%%%%%%%%%%%%%%%
\section{SAMPLE SCENARIOS FOR USING SIMFAAS} \label{sec:sample-use}

\noindent
In this section, we will go through a few sample use cases for the
serverless platform simulator presented in this work. For more details,
a comprehensive list of examples can be found in the project Github
repository\footnote{
\url{https://github.com/pacslab/SimFaaS/tree/master/examples}
}.

\subsection{Steady-State Analysis}
In this example, we use the SimFaaS simulator to calculate the steady-state
properties of a given workload
in scale-per-request serverless computing platforms.
In SimFaaS, the
workload is only characterized by arrival rate,
service time (warm start response time), 
and the
provisioning time (the amount of time to have a cold start
instance get ready to serve the request),
which are easily accessible through experimentation and
any monitoring dashboard. The only information needed to
characterize the serverless computing platform is the expiration
threshold, which is the amount of time it takes for the platform
to expire and recycle the resources of an instance after it has
finished processing its last request. This value is usually constant
and the same for all users of the serverless computing platform.
To run a simple simulation, we can leverage the \textit{ServerlessSimulator} 
class and run the simulation long enough to minimize the transient effect and let the system achieve the steady-state.

\begin{table}[h]
% to adjust the height of each cell
\renewcommand{\arraystretch}{1.3}
\caption{An example simulation input and selected output parameters. The output parameters are signified with a leading star (*).}\label{tab:steady-state-example} \centering
\centering
\rowcolors{2}{white}{gray!25}
\begin{tabular}{c|c}
  \hline
  \textbf{Parameter} & \textbf{Value} \\
  \hline
  Arrival Rate & 0.9 req/s\\
  \hline
  Warm Service Time & 1.991 s\\
  \hline
  Cold Service Time & 2.244 s\\
  \hline
  Expiration Threshold & 10 min\\
  \hline
  Simulation Time & $10^6$ s\\
  \hline
  Skip Initial Time & 100 s\\
  \hline
  *Cold Start Probability & 0.14 \%\\
  \hline
  *Rejection Probability & 0 \%\\
  \hline
  *Average Instance Lifespan & 6307.7389 s\\
  \hline
  *Average Server Count & 7.6795\\
  \hline
  *Average Running Servers & 1.7902\\
  \hline
  *Average Idle Count & 5.8893\\
  \hline
\end{tabular}
\end{table}

\Cref{tab:steady-state-example} shows a set of example simulation parameters
with the default exponential distribution both for the arrival and service time processes.
Note that instead of using exponential distributions, the user can pass a random generator function 
with a custom distribution to achieve more accurate results for specific applications. As can be
seen, the system can produce QoS-related parameters like the probability of cold start or
rejection of the request for a given arrival rate, which can help the application developer
analyze and find the limits of the system. In addition, the application developer can also
use the average number of running servers as an important measure for the cost of their service
that can be used for setting different configurations of services that the function relies on, e.g., the database
concurrent connection capacity~\cite{awsdynamocap}. Besides, information like the average server count can produce an
estimate for the infrastructure cost incurred by the serverless provider. The serverless provider
can use SimFaaS as a tool to analyze the possible effect of changing parameters like the expiration
threshold on their incurred cost and QoS for different scenarios.

\begin{figure}[!htbp]
\centering
\includegraphics[width=\linewidth]{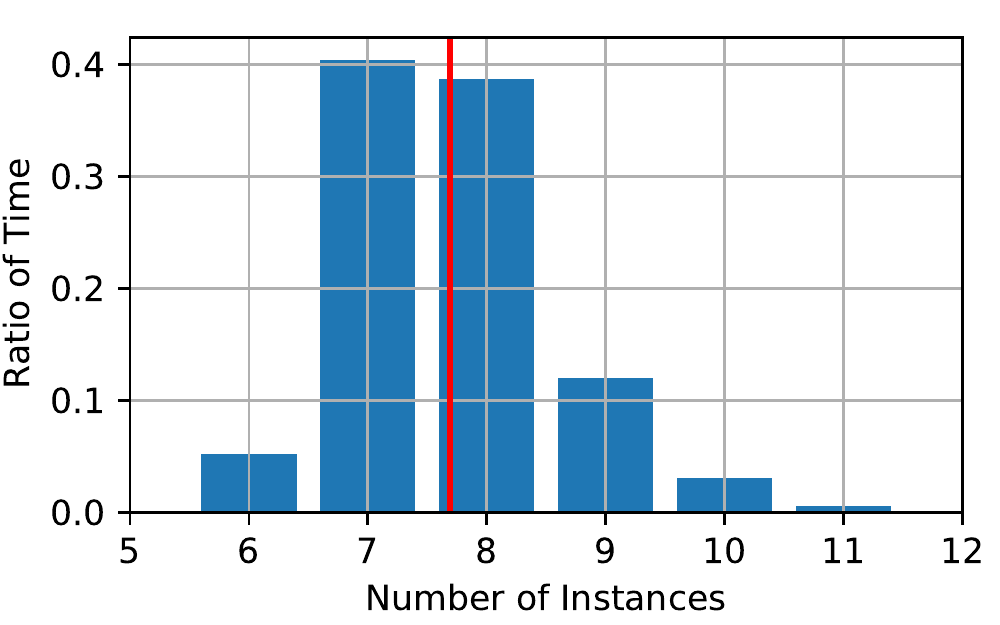}
\caption{The instance count distribution of the simulated process
throughout time. The y-axis represents the portion of time in the
simulation with a specific number of instances.}
\label{fig:example-instcount}
\end{figure}

\begin{figure}[!htbp]
\centering
\includegraphics[width=\linewidth]{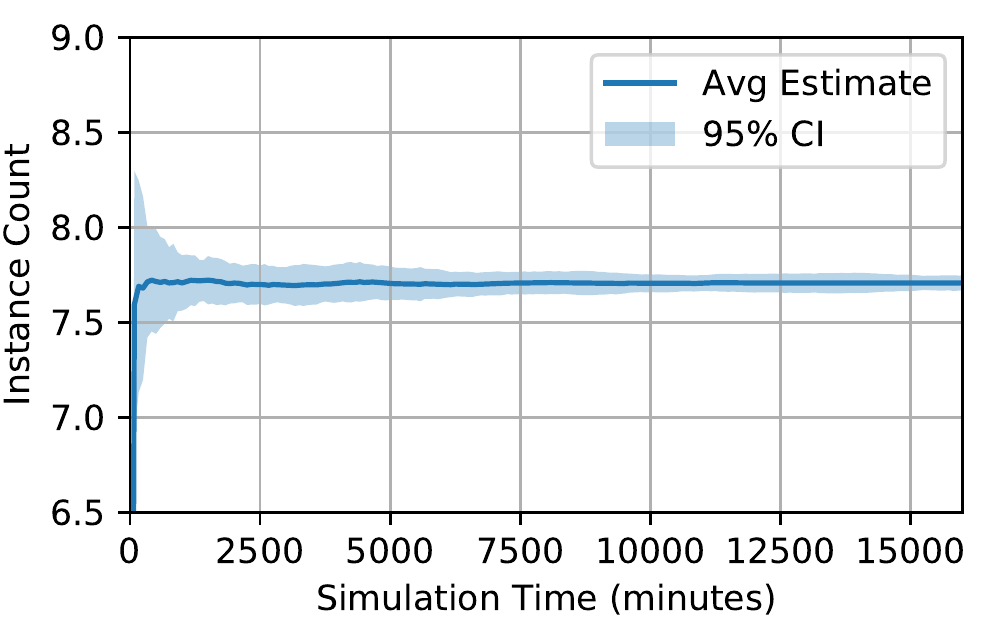}
\caption{The estimated average instance count over time in 10 simulations. The solid line shows the average of simulations and the shaded area shows the 95\% Confidence Interval (CI).}
\label{fig:example-instcount-over-time}
\end{figure}

Another way the proposed simulator can be leveraged is for extracting
information about the system that is not visible to software engineers and developers in public
serverless computing platforms like AWS Lambda or Google Cloud Functions.
This information could facilitate research for predicting cost, performance,
database configurations or other related parameters.
For example, we can find out the distribution
of instance counts in the system throughout time in the simulated
platform for input parameters shown in \Cref{tab:steady-state-example}, 
which is shown in \Cref{fig:example-instcount}. This information can help
researchers develop performance models based on internal states of the 
system with very good accuracy, which is otherwise not possible in
public serverless offerings.
To further analyze the reproducibility of our instance count estimation using the parameters in 
\Cref{tab:steady-state-example}, we ran 10 independent simulations
and generated our estimation of average instance count over time for each run.
\Cref{fig:example-instcount-over-time} shows the average and 95\% confidence interval of our estimation
over those runs. As can be seen, our estimation converges, showing less than 1\%
deviation from the mean in the 95\% confidence interval.

\subsection{Transient Analysis}

Although the steady-state analysis of the serverless computing platform's performance
can give us the long-term quality of service metrics, the application developer or the
serverless provider might be interested in the platform's transient behaviour.
A transient analysis simulation can provide insight into the immediate future, facilitating
time-bound performance guarantees. Besides, it can help serverless providers ensure the
short-term quality of service when trying new designs.

Previous efforts have been made to develop performance models able to provide transient
analysis of serverless computing platforms~\cite{mahmoudi2020tempperf}. However, there
are inherent limitations to such performance models, like the absence of batch arrival modelling
and being limited to Markovian processes. SimFaaS doesn't
have such limitations and can help both application developers and serverless providers gain
insight into the transient aspect of the performance of serverless computing platforms.

\subsection{What-If Analysis}

Due to the inherent highly-dynamic infrastructure of serverless computing platforms,
there are very few tools from the performance engineering methodologies and analysis that
can be used in the emerging serverless technologies. Because of this inherent lack of
tools and resources, serverless computing platforms were forced to use trial and error
through implementation to analyze their new designs for making performance and efficiency
improvements. There have been previous studies that proposed analytical performance models
for serverless computing platforms \cite{mahmoudi2020tccserverless,mahmoudi2020tempperf},
but these methods have limitations like only supporting Markovian processes, limiting their
applicability in a number of scenarios.

\begin{figure}[!htbp]
\centering
\includegraphics[width=\linewidth]{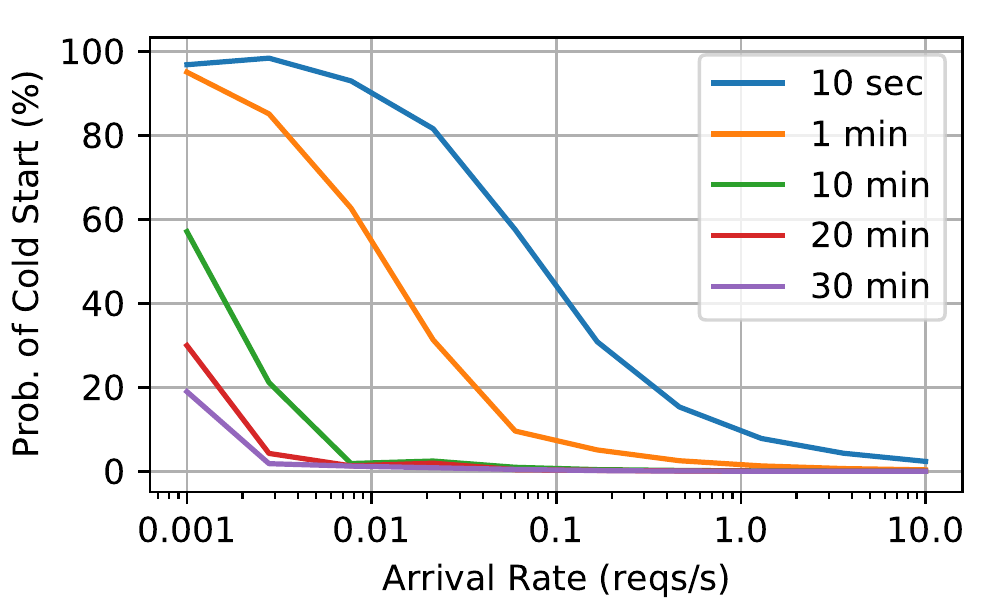}
\caption{Cold start probability against the arrival rate for different
values of the expiration threshold for the workload specified in \Cref{tab:steady-state-example}. SimFaaS can ease
the process of conducting experiments with several configurations
to find the best performing one.}
\label{fig:05-example-what-if}
\end{figure}

One major benefit of having an accurate serverless platform simulator is the
ability to perform what-if analysis on different configurations and
find the best-performing settings for a given workload. Implementation
and experimentation to gather similar data is both time-consuming
and costly, while using the proposed simulator makes
the data collection much faster and easier.
\Cref{fig:05-example-what-if} shows an example of such an analysis for
different values for the \textit{expiration threshold} in the system.
Different workloads running on serverless computing platforms might
have different performance/cost criteria. Using what-if analysis
powered by an accurate performance model, one could optimize the
configuration for each unique workload.
Similar examples can be found in the project examples.

\subsection{Cost Calculation}

Performing cost prediction under different loads in cloud computing is generally a very challenging
task. These challenges tend to be exacerbated in the highly dynamic structure of serverless
computing platforms. Generally, there is a broad range of possible costs for a given serverless
function, including computation, storage, networking, database, or other API-based services like 
machine learning engines or statistical analysis. However, all charges incurred by serverless functions
can be seen as either per-request charges (e.g., external APIs, machine learning, face recognition, network I/O)
or runtime charges billed based on execution time (e.g., memory or computation). Per-request charges can
be calculated using only the average arrival rate. However, runtime charges may differ under different load
intensities due to the difference in cold start probability. Using the proposed
simulator, users can get an estimate on the cold start probability and the average number of running
servers, which are necessary for cost estimation under different load intensities.

In addition to the average running server count, which helps estimate the cost incurred by the 
application developer, the average total server count is linearly proportional to the infrastructure
cost incurred by the serverless provider. Thus, using the proposed simulator, both developer charges
and infrastructure charges incurred by the provider can be estimated under different configurations,
which can help improve the platform by studying the effect of using different configurations or designs
without the need to perform expensive or time-consuming experiments or implementations.

\section{Experimental Evaluation} \label{sec:exp-eval}

\noindent
To show the accuracy of the proposed simulator, we performed extensive
experimentations on AWS Lambda and showed that the results were in
tune with the results from SimFaaS. The experiments
are above one month of running benchmark applications on AWS
Lambda and are openly accessible on 
Github\footnote{
\url{https://github.com/pacslab/serverless-performance-modeling/tree/master/experiments/results}
}.
All of our experiments were executed for a 28-hour window with 10 minutes of warm-up
time in the beginning, during which we do not record any data.
The workload used in this work was based on the work of Wang et al.~\cite{wang2018peeking}
with minor modifications. Our workload is openly available in our Github 
repository\footnote{
\url{https://github.com/pacslab/serverless-performance-modeling}
}. For the purpose
of experimental validation, we used a combination of CPU intensive and I/O intensive workloads.
During the experimentation, we have obtained performance metrics and the other parameters such
as cold/warm start information, instance id, lifespan, etc.

\subsection{Experimental Setup}
In our AWS Lambda deployment, we used the \textit{Python 3.6} runtime
with 128 MB of RAM deployed on the \textit{us-east-1} region
in order to have the lowest possible latency from our client.
Note that the memory configuration won't affect the accuracy of the
simulation as the results depend on the service time distribution,
which captures the effect of changing the memory configuration.
For the client, we used a virtual machine with 8 vCPUs, 16 GB of memory, and 1000 Mbps network connectivity
with single-digit milliseconds latency to AWS servers
hosted on Compute Canada Arbutus cloud\footnote{\url{https://docs.computecanada.ca/wiki/Cloud_resources}}. 
We used Python as the programming language
and the official \textit{boto3} library to communicate with the AWS Lambda API to make the requests
and process the resulting logs for each request. For load-testing and generation of the client requests
based on a Poisson process, we used our in-house workload generation
library\footnote{
\url{https://github.com/pacslab/pacswg}
},
which is openly accessible through PyPi\footnote{
\url{https://pypi.org/project/pacswg}
}.
The result is stored in a CSV file and then processed using Pandas, Numpy, Matplotlib, and Seaborn.
The dataset, parser, and the code for extraction of system parameters and properties are also
publicly available in our analytical model project's Github repository\footnote{
\url{https://github.com/pacslab/serverless-performance-modeling}
}.

\subsection{Parameter Identification}
We need to estimate the system characteristics to be used in our simulator as input parameters. In this section, we discuss
our approach to estimating each of these parameters.

\textbf{Expiration Threshold:} here, our goal is to measure the expiration threshold,
which is the amount of time after
which inactive function instance in the warm pool will be expired and therefore terminated. To measure this parameter,
we created an experiment in which we make requests with increasing inter-arrival times until we
see a cold start meaning that the system has terminated the instance between two consecutive requests.
We performed this experiment on AWS lambda with the starting inter-arrival time of 10 seconds, each time
increasing it by 10 seconds until we see a cold start. In our experiments, AWS lambda instances seemed to expire
an instance exactly after 10 minutes of inactivity (after it has processed its last request). This number
did not change in any of our experiments leading us to assume it is a deterministic value. 
This observation has also been verified in \cite{shilkovaws10mins,shahrad2020serverless}.

\textbf{Average Warm Response Time and Average Cold Response Time:}
with the definitions provided here, warm
response time is the service time of the function, and
cold response time includes both provisioning time and
service time.
To measure the average warm response time and the average cold response 
time, we used the average of response times measured throughout the experiment. 

\subsection{Simulator Results Validation}
In this section, we outline our methodology for measuring the performance metrics of the system,
comparing the results with the predictions of our simulator.

\textbf{Probability of Cold Start:} to measure the probability of cold start, we divide
the number of requests causing a cold start by the total number of requests made during our experiment.
Due to the inherent scarcity of cold starts in most of our experiments,
we observed an increased noise in our measurements for the probability of cold start, which led to increasing the window for data collection to about 28 hours for each sampled point.

\textbf{Mean Number of Instances in the Warm Pool:} to measure the mean number of instances in the warm pool,
we count the number of unique instances that have responded to the client's requests in the past 10 minutes.
We use a unique identifier for each function instance to keep track of their life cycle, as obtained in~\cite{wang2018peeking}.

\textbf{Mean Number of Running Instances:} we calculate this metric by observing the system every ten seconds, counting the number of in-flight requests in the system, taking the average as our estimate.

\textbf{Mean Number of Idle Instances:} this can be measured as the difference between the total average number of instances in the warm pool and the number of instances busy running the requests.

\textbf{Average Wasted Capacity:} for this metric, we define the utilized capacity as the
ratio of the number of running instances over all instances in the warm pool. Using this
definition, the ratio of idle instances over all instances in the warm pool is the
wasted portion of capacity provisioned for our workload.
Note that this value is very important to the provider as it measures the
ratio of the utilized capacity (billed for the application developer) over the deployed capacity (reflecting the infrastructure cost).

\subsection{Experimental Results}
\label{sec:exp-results}

\begin{figure}[!ht]
\centering
\includegraphics[width=\linewidth]{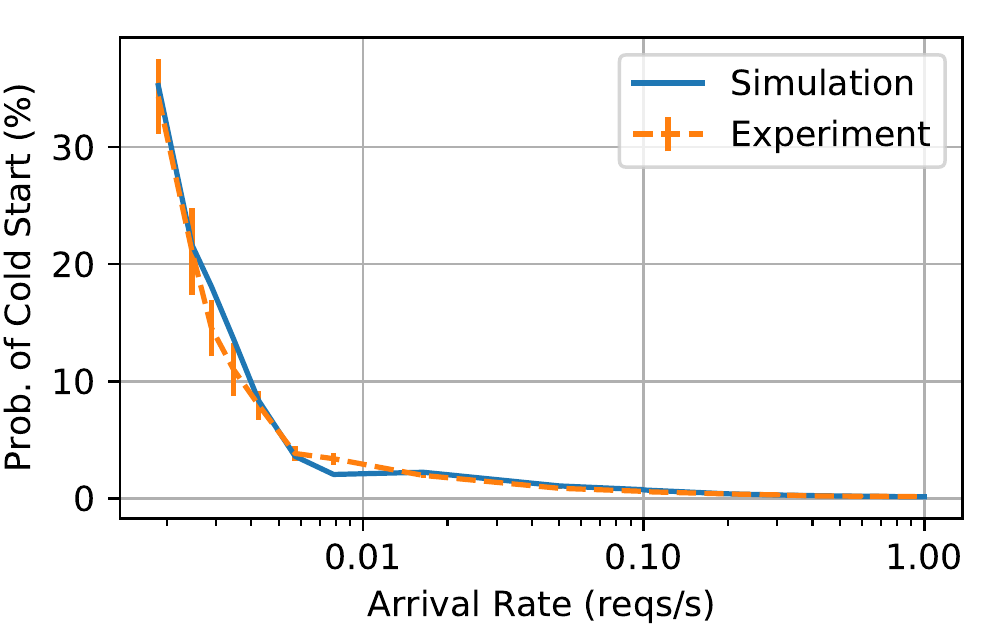}
\caption{Probability of cold start extracted from simulation compared with real-world experimentations on AWS Lambda.
}
\label{fig:04-experiment-comp}
\end{figure}

\begin{figure}[!ht]
\centering
\includegraphics[width=\linewidth]{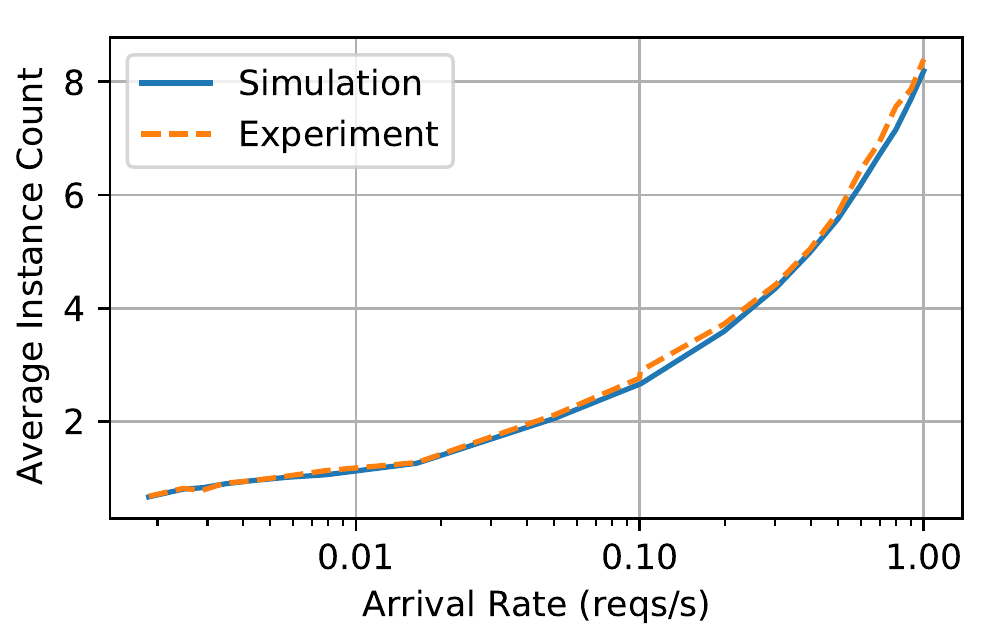}
\caption{The average number of instances extracted from simulation compared with real-world experimentations on AWS Lambda.}
\label{fig:04-experiment-instcount}
\end{figure}

\begin{figure}[!ht]
\centering
\includegraphics[width=\linewidth]{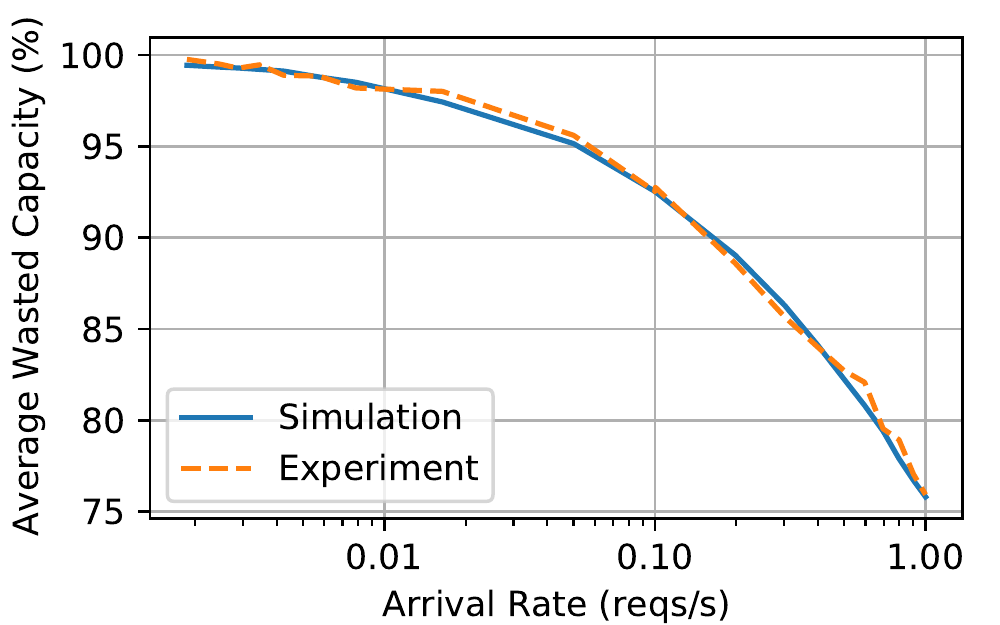}
\caption{Average wasted resources extracted from simulation compared with real-world experimentations on AWS Lambda.}
\label{fig:04-experiment-wated}
\end{figure}

\Cref{fig:04-experiment-comp} shows the probability of cold start for
different arrival rates extracted from the simulation compared with
real-world results. As can be seen, the results match the performance
metrics extracted from experimentations. The results show an average
error of $12.75\%$, while the standard error of the underlying process for the experiments
over $10$ runs totalling $28$ hours is $10.14\%$ showing the accuracy
of the results obtained from the simulation.
\Cref{fig:04-experiment-instcount,fig:04-experiment-wated}
show the average number of instances and the average wasted
capacity (in \textit{idle} state) for the simulation and
experiments with Mean Absolute Percentage Error (MAPE) of
$3.43\%$ and $0.17\%$, respectively.

\section{RELATED WORK} \label{sec:related-work}

\noindent
Many recent works in the area of serverless computing platforms have focused on studying and finding ways to improve the
performance of serverless computing platforms~\cite{van2018addressing,manner2018cold,manner2019towards,boza2017reserved,abad2018package,jeon2019cloudsim}. However, to the best of the authors' knowledge, none have
been able to predict or simulate comprehensive performance or quality of metrics characteristics for a given workload.
In this section, we will go through the most related recent works in the prediction of the performance or quality
of service for serverless computing platforms.

Some of the previous studies have focused on developing comprehensive performance models for steady-state and transient analysis
of a given workload~\cite{mahmoudi2020tccserverless,mahmoudi2020tempperf}. However, the proposed models impose
several limitations on the arrival and service time processes and cannot handle batch arrivals. These limitations
render the developed performance model unusable for many types of workloads, especially for batch and analytics
workloads. The serverless performance simulator proposed in this work can handle any type of arrival or service time
process and can be adapted to future emerging serverless computing management models with less manual effort.

% Investigating Performance and Cost in Function-as-a-Service Platforms},
In~\cite{bortolini2019investigating}, Bortolini et al. used experimentations on different configurations and serverless providers in order to find the most important factors influencing the performance and cost of current serverless platforms. In their study, they found low predictability of cost as one of the most important drawbacks of serverless computing platforms. Using simulators like what we proposed in this work can help improve the predictability of the cost of a given workload under different load intensities.
% Serverless Computing: One Step Forward, Two Steps Back
Hellerstein et al.~\cite{hellerstein2018serverless}
addressed the main shortcomings present in the first-generation
serverless computing platforms and the anti-patterns
present in them. They showed that current implementations are
restricting distributed programming and cloud computing innovations.
% A SPEC RG Cloud Group’s Vision on the Performance Challenges of FaaS Cloud Architectures
Eyk et al.~\cite{van2018spec} found the most important issues surrounding the widespread 
adoption of FaaS to be sizeable overheads, unreliable performance, and new forms of 
cost-performance trade-off. In their work, they identified six performance-related challenges 
for the domain of serverless computing and proposed a roadmap for alleviating these challenges.
Several of the aforementioned shortcomings of serverless computing platforms can be mitigated by predicting
the cost-performance trade-off using serverless simulators.

% Addressing Performance Challenges in Serverless Computing
Eyk et al.~\cite{van2018addressing} investigated the performance challenges in current
serverless computing platforms. They found the most important challenges in the adoption
of FaaS to be the remarkable computational overhead, unreliable performance, and absence of benchmarks.
The introduction of an accurate simulator for function-as-a-service offerings could overcome some of these
shortcomings.
% Cold Start Influencing Factors in Function as a Service
Manner et al.~\cite{manner2018cold} investigated the factors influencing the cold start performance of serverless computing platforms. Their experiments on AWS Lambda and Azure Functions show that factors like the programming language, deployment package size, and memory settings affect the performance on serverless computing platforms.
% ================== Serverless Simulators ==================
In a later study,
Manner et al.~\cite{manner2019towards} describe the importance of an accurate simulator for Function-as-a-Service (FaaS) products. They mention how scaling, cold starts, function configurations, dependent services, network latency, and other important configurations influence cost-performance trade-off. In their work, they propose a simulation framework for a cost and performance simulator for serverless computing platforms. In this platform, they suggest extracted mean values from experiments as inputs to the performance model in order to calculate different properties.

Boza et al.~\cite{boza2017reserved} introduced a model-based simulation for cloud budget planning. In their work, they perform cost simulation for using reserved VMs, on-demand VMs, bid-based VMs, and serverless computing for a similar computing task. In their work, the serverless computing simulation is overly simplistic for performance modelling researchers and lacks several important details. In this work, we focus solely on performance simulation of serverless computing platforms, but with more details in mind, which seems necessary for the simulator to be leveraged by the performance research community, application developers, and serverless providers.

Abad et al.~\cite{abad2018package} mainly considered the problem of scheduling small cloud functions on serverless computing platforms. As a part of their evaluations, they implemented a SimPy-based simulation for their proposed scheduling method. Although this work shows promise of rather accurate serverless computing simulations, their focus is on scheduling tasks while ignoring several details of interest for performance modelling. In this work, we strive to fill this gap by providing the performance modelling research with the proper tooling necessary for high-fidelity performance models of serverless computing platforms.
Jeon et al.~\cite{jeon2019cloudsim} introduced a CloudSim extension focused on Distributed Function-as-a-Service (DFaaS) on edge devices. Although the DFaaS systems hold a great promise for the future of serverless computing, it doesn't allow simulation for the mainstream serverless computing platforms.

\section{CONCLUSION AND FUTURE WORK} \label{sec:conc}

\noindent
In this work, we presented SimFaaS, a simulator for modern
serverless computing platforms with sufficient details to yield very
accurate results. We introduced a range of tools available for
performance modelling researchers giving them insights and details
into several internal properties that are not visible for users
in public serverless computing platforms. We reviewed some of the
possible use cases of the proposed simulator and showed its accuracy
through comparison with real-world traces gathered from running benchmark
applications on AWS Lambda.

SimFaaS enables performance modelling researchers with a tool allowing them
to develop accurate performance models using
the internal state of the system, which cannot be monitored on public
serverless computing platforms.
Using SimFaaS, both serverless providers and
application developers can predict the quality of service,
expected infrastructure and incurred cost, amount of 
wasted resources, and energy consumption without performing
lengthy and expensive experimentations.
The benefits of using SimFaaS for serverless computing platform
providers could be two-fold:
1) They can examine new designs, developments, and deployments in their platforms by initially validating new ideas on SimFaaS, which will be significantly cheaper in terms of time and cost compared to actual prototyping;
2) They can provide users with fine-grain control over the cost-performance
trade-off by modifying the platform parameters (e.g., \textit{expiration threshold}).
This is mainly due to the fact that there is no universal optimal point in the
cost-performance trade-off for all workloads. By making accurate
predictions, a serverless provider can better optimize their resource
usage while improving the application developers' experience and consequently the end-users.

As future work, we plan to extend SimFaaS to include new
generations of serverless computing, in addition to adding
several billing schemas in order to predict the cost of
running workloads on the serverless platform.
We are also aiming to maintain SimFaaS by the members
of 
PACS
Lab\footnote{
\url{https://pacs.eecs.yorku.ca}
}
to add new features offered by serverless providers.

\section*{Acknowledgements}

\noindent
This research was enabled in part by support provided by Sharcnet (www.sharcnet.ca) and Compute Canada (www.computecanada.ca).
We would also like to thank Amazon for supporting this research by providing us with the education credit to
access the Amazon Web Services (AWS).

\bibliographystyle{apalike}
{\small
\bibliography{bibliography}}

\begin{thebibliography}{}

\bibitem[Abad et~al., 2018]{abad2018package}
Abad, C.~L., Boza, E.~F., and Van~Eyk, E. (2018).
\newblock Package-aware scheduling of faas functions.
\newblock In {\em Companion of the 2018 ACM/SPEC International Conference on
  Performance Engineering}, pages 101--106.

\bibitem[{Amazon Web Services Inc.}, 2020]{awsdynamocap}
{Amazon Web Services Inc.} (2020).
\newblock {Amazon DynamoDB Read/Write Capacity Mode}.
\newblock
  \url{https://docs.aws.amazon.com/amazondynamodb/latest/developerguide/HowItWorks.ReadWriteCapacityMode.html}.
\newblock Last accessed 2020-11-20.

\bibitem[Bermbach et~al., 2020]{bermbachusing}
Bermbach, D., Karakaya, A.~S., and Buchholz, S. (2020).
\newblock Using application knowledge to reduce cold starts in faas services.
\newblock In {\em Proceedings of the 35th ACM/SIGAPP Symposium on Applied
  Computing}.

\bibitem[Bortolini and Obelheiro, 2019]{bortolini2019investigating}
Bortolini, D. and Obelheiro, R.~R. (2019).
\newblock Investigating performance and cost in function-as-a-service
  platforms.
\newblock In {\em International Conference on P2P, Parallel, Grid, Cloud and
  Internet Computing}, pages 174--185. Springer.

\bibitem[Boza et~al., 2017]{boza2017reserved}
Boza, E.~F., Abad, C.~L., Villavicencio, M., Quimba, S., and Plaza, J.~A.
  (2017).
\newblock Reserved, on demand or serverless: Model-based simulations for cloud
  budget planning.
\newblock In {\em 2017 IEEE Second Ecuador Technical Chapters Meeting (ETCM)},
  pages 1--6. IEEE.

\bibitem[Figiela et~al., 2018]{figiela2018performance}
Figiela, K., Gajek, A., Zima, A., Obrok, B., and Malawski, M. (2018).
\newblock Performance evaluation of heterogeneous cloud functions.
\newblock {\em Concurrency and Computation: Practice and Experience},
  30(23):e4792.

\bibitem[{Google Cloud Platform Inc.}, 2020]{gcpconcurrency}
{Google Cloud Platform Inc.} (2020).
\newblock Concurrency.
\newblock \url{https://cloud.google.com/run/docs/about-concurrency}.
\newblock Last accessed 2020-02-13.

\bibitem[Hellerstein et~al., 2018]{hellerstein2018serverless}
Hellerstein, J.~M., Faleiro, J., Gonzalez, J.~E., Schleier-Smith, J.,
  Sreekanti, V., Tumanov, A., and Wu, C. (2018).
\newblock Serverless computing: One step forward, two steps back.
\newblock {\em arXiv preprint arXiv:1812.03651}.

\bibitem[Jeon et~al., 2019]{jeon2019cloudsim}
Jeon, H., Cho, C., Shin, S., and Yoon, S. (2019).
\newblock A cloudsim-extension for simulating distributed
  functions-as-a-service.
\newblock In {\em 2019 20th International Conference on Parallel and
  Distributed Computing, Applications and Technologies (PDCAT)}, pages
  386--391. IEEE.

\bibitem[Lin and Glikson, 2019]{lin2019mitigating}
Lin, P.-M. and Glikson, A. (2019).
\newblock Mitigating cold starts in serverless platforms: A pool-based
  approach.
\newblock {\em arXiv preprint arXiv:1903.12221}.

\bibitem[Lloyd et~al., 2018]{lloyd2018serverless}
Lloyd, W., Ramesh, S., Chinthalapati, S., Ly, L., and Pallickara, S. (2018).
\newblock Serverless computing: An investigation of factors influencing
  microservice performance.
\newblock In {\em 2018 IEEE International Conference on Cloud Engineering
  (IC2E)}, pages 159--169. IEEE.

\bibitem[Mahmoudi and Khazaei, 2020a]{mahmoudi2020tccserverless}
Mahmoudi, N. and Khazaei, H. (2020a).
\newblock {Performance Modeling of Serverless Computing Platforms}.
\newblock {\em IEEE Transactions on Cloud Computing}, pages 1--15.

\bibitem[Mahmoudi and Khazaei, 2020b]{mahmoudi2020tempperf}
Mahmoudi, N. and Khazaei, H. (2020b).
\newblock {Temporal Performance Modelling of Serverless Computing Platforms}.
\newblock In {\em {Proceedings of the 6th International Workshop on Serverless
  Computing}}, WOSC '20, pages 1--6. Association for Computing Machinery.

\bibitem[Mahmoudi et~al., 2019]{mahmoudi2019optimizing}
Mahmoudi, N., Lin, C., Khazaei, H., and Litoiu, M. (2019).
\newblock Optimizing serverless computing: introducing an adaptive function
  placement algorithm.
\newblock In {\em Proceedings of the 29th Annual International Conference on
  Computer Science and Software Engineering}, pages 203--213.

\bibitem[Manner, 2019]{manner2019towards}
Manner, J. (2019).
\newblock Towards performance and cost simulation in function as a service.
\newblock {\em Proc. ZEUS (accepted)}.

\bibitem[Manner et~al., 2018]{manner2018cold}
Manner, J., Endre{\ss}, M., Heckel, T., and Wirtz, G. (2018).
\newblock Cold start influencing factors in function as a service.
\newblock In {\em 2018 IEEE/ACM International Conference on Utility and Cloud
  Computing Companion (UCC Companion)}, pages 181--188. IEEE.

\bibitem[McGrath and Brenner, 2017]{mcgrath2017serverless}
McGrath, G. and Brenner, P.~R. (2017).
\newblock Serverless computing: Design, implementation, and performance.
\newblock In {\em 2017 IEEE 37th International Conference on Distributed
  Computing Systems Workshops (ICDCSW)}, pages 405--410. IEEE.

\bibitem[{Mikhail Shilkov}, 2020]{shilkovaws10mins}
{Mikhail Shilkov} (2020).
\newblock Cold starts in aws lambda.
\newblock \url{https://mikhail.io/serverless/coldstarts/aws/}.
\newblock Last accessed 2020-03-18.

\bibitem[Shahrad et~al., 2020]{shahrad2020serverless}
Shahrad, M., Fonseca, R., Goiri, {\'I}., Chaudhry, G., Batum, P., Cooke, J.,
  Laureano, E., Tresness, C., Russinovich, M., and Bianchini, R. (2020).
\newblock Serverless in the wild: Characterizing and optimizing the serverless
  workload at a large cloud provider.
\newblock {\em arXiv preprint arXiv:2003.03423}.

\bibitem[van Eyk and Iosup, 2018]{van2018addressing}
van Eyk, E. and Iosup, A. (2018).
\newblock Addressing performance challenges in serverless computing.
\newblock In {\em Proc. ICT. OPEN}.

\bibitem[Van~Eyk et~al., 2018]{van2018spec}
Van~Eyk, E., Iosup, A., Abad, C.~L., Grohmann, J., and Eismann, S. (2018).
\newblock A spec rg cloud group's vision on the performance challenges of faas
  cloud architectures.
\newblock In {\em Companion of the 2018 ACM/SPEC International Conference on
  Performance Engineering}, pages 21--24. ACM.

\bibitem[Wang et~al., 2018]{wang2018peeking}
Wang, L., Li, M., Zhang, Y., Ristenpart, T., and Swift, M. (2018).
\newblock Peeking behind the curtains of serverless platforms.
\newblock In {\em 2018 USENIX Annual Technical Conference (USENIX ATC 18)},
  pages 133--146.

\end{thebibliography}

\end{document}